\documentstyle[psfig,prb,aps]{revtex}
\begin{document}
\title{Insulating gap in FeO: correlations and covalency}
\author{I. I. Mazin$^{a,b}$ and V.I. Anisimov$^{a,c}$}
\address{$^a$Max-Planck-Institut f\"{u}r Festk\"{o}rperforschung, Stuttgart D-70569,\\
Germany\\
$^b$Geophysical Laboratory and Center for High-Pressure Research,\\
CarnegieInstitution of Washington, 5251 Broad Branch Rd. N.W., Washington,\\
D.C. 20005-1305 U.S.A.\\
$^c$Institute for Metal Physics, Ekaterinburg, Russia}

\twocolumn[\hsize\textwidth\columnwidth\hsize\csname@twocolumnfalse\endcsname
\maketitle

\begin{abstract}
We report calculations of the electronic structure of FeO in the LDA and
LDA+U approximation with and without rhombohedral distortion. In both cases
LDA renders an antiferromagnetic metal, and LDA+U opens a Hubbard gap. 
However, the character of the gap is qualitatively different in the
two structure, and the difference can be traced down to underlying
LDA bnad structure.  An analysis of the calculations gives a
new insight on the origin of the insulating gap in 3d monoxides and on the
role of the {\bf k-}dependency of U, missing in the contemporary LDA+U
method.
\end{abstract}

\draft
]

It is a well-known fact that conventional band structure calculations
incorrectly give metallic ground state for the intermediate 3d transition
metal monoxides, CoO and FeO. Before the high-$T_c$ cuprates entered the
scene, this had been often considered as the most notable failure of the
Local Density Approximation (LDA). In the last decade a number of extensions
of the density-functional theory were suggested, which, in different
manners, led to insulating ground states for this compounds. Most successful
were various flavors of the Self-Interaction Corrected (SIC) LDA\cite{sic},
LDA+U\cite{LDA+U}, and the Orbital Polarization Correction in a crystal
field basis\cite{norman}. Interestingly, apart from the gap itself, LDA
appears to do as good, and sometimes a better job, than these sophisticated
extensions, especially when Generalized Gradient Correction to the
conventional LDA is taken into account. Structural properties are reproduced
very well, including rhombohedral distortion in FeO\cite{coo}, and its
increase with pressure\cite{FeO}. Moreover, angular resolved photoemission
renders the bands more similar to the LDA bands, than to those in other
calculational scheme (except for the narrow range near the gap)\cite{zx}.
The magnitude of the magnetic moment, which is often believed to be the
first indication of an LDA failure, is nearly exact for the spin moment in
CoO (2.4 $\mu _B$) and in FeO (3.5 $\mu _B$, assuming no orbital moment). On
the contrary, the gap-improving calculation tend to overestimate the Fe
moment, especially when the orbital moment is included (i.e., there is a
tendency \cite{norman} to underestimated the crystal field quenching of the
orbital moment). It seems, then, that the main problem in the LDA is purely
spectroscopical (quite in the spirit of the density functional theory),
namely non-existence of the gap. It is worth noting that an important common
feature of the LDA and the corrected schemes mentioned above is substantial
width of the metal $d$-bands (for instance, in LDA the width of Fe $t_{2g}$
band at normal pressure is about 1.4 eV). This should be contrasted with the
popular analysis of the electronic structure of 3$d$-oxides in terms of
separate levels of a width less than Jan-Teller energy and spin-orbital
coupling\cite{Good+Dan}.

The fact that LDA gaps are always too small is well understood.
Mathematically it appears as the density-derivative discontinuity of the
Kohn-Sham potential in the DFT. The physics of this discontinuity may be
different, but for transition metal oxides (NiO, MnO) it is usually
associated with the Mott-Hubbard repulsion\cite{norman}. In view of this, it
has always been much more disturbing to have wrongly a metallic behavior in
LDA calculation, than just to have a wrong gap. As formulated by Norman \cite
{norman}, ``one would like to obtain a gap at the level of a density
functional calculation (no matter how small) so as to define the
Mott-Hubbard correction in an unambiguous fashion''. Moreover, even in a
case when LDA does not give a gap, but gives reasonable band structure
except for the immediate vicinity of the Fermi level, and correctly
describes delicate features of the ground state, like magnetoelastic
interactions, it is desirable to have a correction scheme which does not
destroy the LDA bands completely, but rather corrects them in a systematic
manner. Unfortunately none of the schemes above acts in such a way. In this
paper we shall analyze the results of the rotationally invariant LDA+U \cite
{LA} calculations for FeO in more details than  is usually done, and
compare them with the standard LDA calculations, paying particular attention
to the process of the LDA+U gap opening in cubic and rhombohedral structure.
We will see that the ground state in the LDA+U approach 
is intimately related with
the underlying LDA band structure, although LDA+U cannot fully account for
the bands hybridization effects, which seem to be quite important here. We
will argue that none of the existing ``corrected LDA'' schemes (nor the
straight LDA) correctly describes the insulating ground state in FeO and
similar compounds. On the other hand, non-local schemes similar to the GW
approximation may be able to provide a qualitatively correct description. It
is worth noting that none of the existing  ``extended-LDA'' calculations has
taken  into account such an important factor as the distortion from the
ideal cubic NaCl-type structure, which is associated in 3$d$-monoxides with
the onset of magnetic ordering. 

Electronic structure of an isolated Fe$^{++}$ ion in a cubic field is
described in the high-spin state by the following scheme\cite{griffith}: The
spin-up $d$-states are all filled, and separated from the spin-down states
by the exchange splitting $E_{ex}$. The partially occupied spin-down states
are split by the crystal field, so that $\epsilon (e_g)-\epsilon
(t_{2g})=\Delta \ll E_{ex}.$
There is one electron in the $t_{2g\downarrow }$ state, which is triply
degenerate. Antiferromagnetic FeO has close-packed (111) planes of Fe
ions with the same spin, which alternate with the similar planes with the
opposite spin, thus lowering the symmetry to rhombohedral one. In the
rhombohedral field, there is one linear combination of the $t_{2g}$ states,
namely $A_{1g}=(xy+yz+zx)/\sqrt{3},$ which can also be written as $%
3Z^2-r^2, $ where $Z$ is parallel to [111]
 (here and below we shall use lower case symbols for the states
classification in cubic symmetry and upper case symbols for the
 rhombohedral nomenclature). The four
other states have the same symmetry $E_g$ (in the rhombohedral nomenclature),
but when the deviation from the
cubic symmetry is small one can speak about the upper two levels, $%
E_g^{^{\prime \prime }},$ and the lower two levels, $E_g^{^{\prime }},$
which are close to the $A_{1g}$
 also originate from the cubic $t_{2g}$ states,
 and are separated from $E_g^{^{\prime \prime }}$
by approximately $\Delta $. In the LDA calculation, when the crystal
symmetry is still cubic, although the magnetic ordering is rhombohedral, the
splitting between  $E_g^{^{\prime }}$ and $A_{1g}$ is small, much smaller
than their bandwidths, so that they merge into one band (corresponding
to the cubic $t_{2g}$ band), which is
necessarily metallic. Note that although the magnetism in this system
appears due to the indirect exchange and is determined by the O-Fe $pd$
interaction, the
width of the $t_{2g}$ band is mainly due to the Fe-Fe $dd\sigma $ overlap.
Fe-O hopping for this band is mainly $pd\pi $ and weak. The situation in CoO
with two $t_{2g}$ electrons is very similar.

There were several successful attempts to obtain an insulating state in FeO 
\cite{sic,LDA+U,norman}. Interestingly, all these approaches give insulating
gap in a fair agreement with the experiment, but all because of different
reasons. SIC calculations \cite{sic} favor $d-$bands in pure orbital states,
that is, undermining the role of crystal field and quenching of orbital
moment. The effect on the occupied $d$-states is thus extremely strong (of
the order of 1 Ry) and all four oxides come out as pure charge transfer
insulators. This is in contradiction with the general experimental indication 
that the character of the band gap changes from predominantly Mott-Hubbard
to predominantly charge-transfer when going from MnO to NiO. Large orbital
moment is obtained for FeO, in contradiction with the experiment (an
argument is usually made that the experimental number may be incorrect
because poor sample quality). It is also worth noting that the SIC formalism
was initially invented as a remedy of the LDA in the direction of the exact,
self-interaction free density functional. It is hard to imagine, however,
that the exact density functional theory with its orbital-independent
one-electron potential can unquench the orbital moment.

The orbital-corrected functional used by
 Norman \cite{norman} has a similar problem.
The correct many-body solution for an isolated ion with unfilled $t_{2g}$
shell in a cubic field\cite{griffith} has an energy contribution
proportional to the total angular momentum. This term, however, does not
favor a specific direction of the momentum, that is, does not include the
projections. The only interaction which does unquench the orbital momentum
is spin-orbit, and it is relatively weak. Norman's correction has the same
functional form as the exact quantum chemical expression\cite{griffith}, but
substitutes the total momentum by its projection on the quantization axis.
This orbital moment projection dependent term acts in a way similar to  the
spin-orbit coupling, but with a much larger magnitude (of the order of
exchange splitting $J$). Thus in FeO the occupied spin-down $d$-band is too
close to a pure $m=1$ state, that is, $xz+iyz.$
\begin{figure}
\centerline{\psfig{file=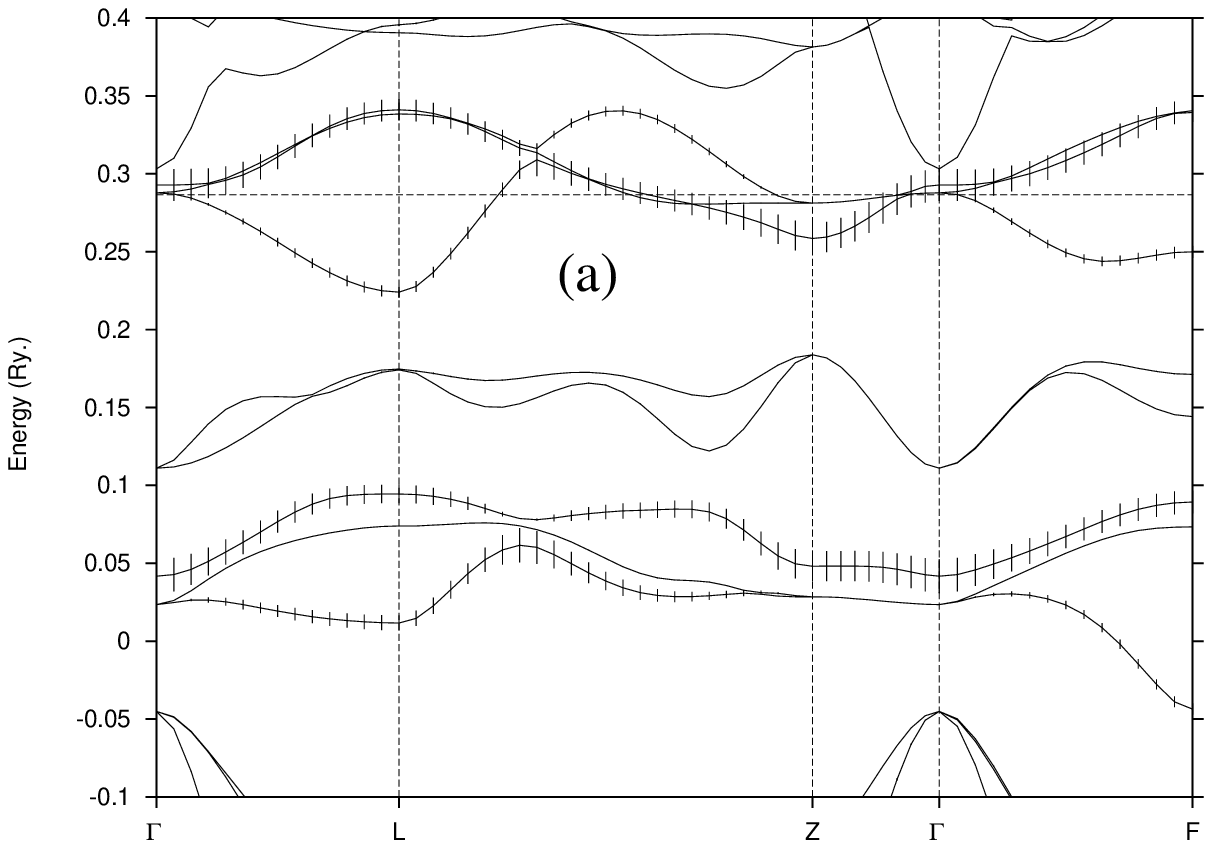,width=0.95\linewidth}} %
\centerline{\psfig{file=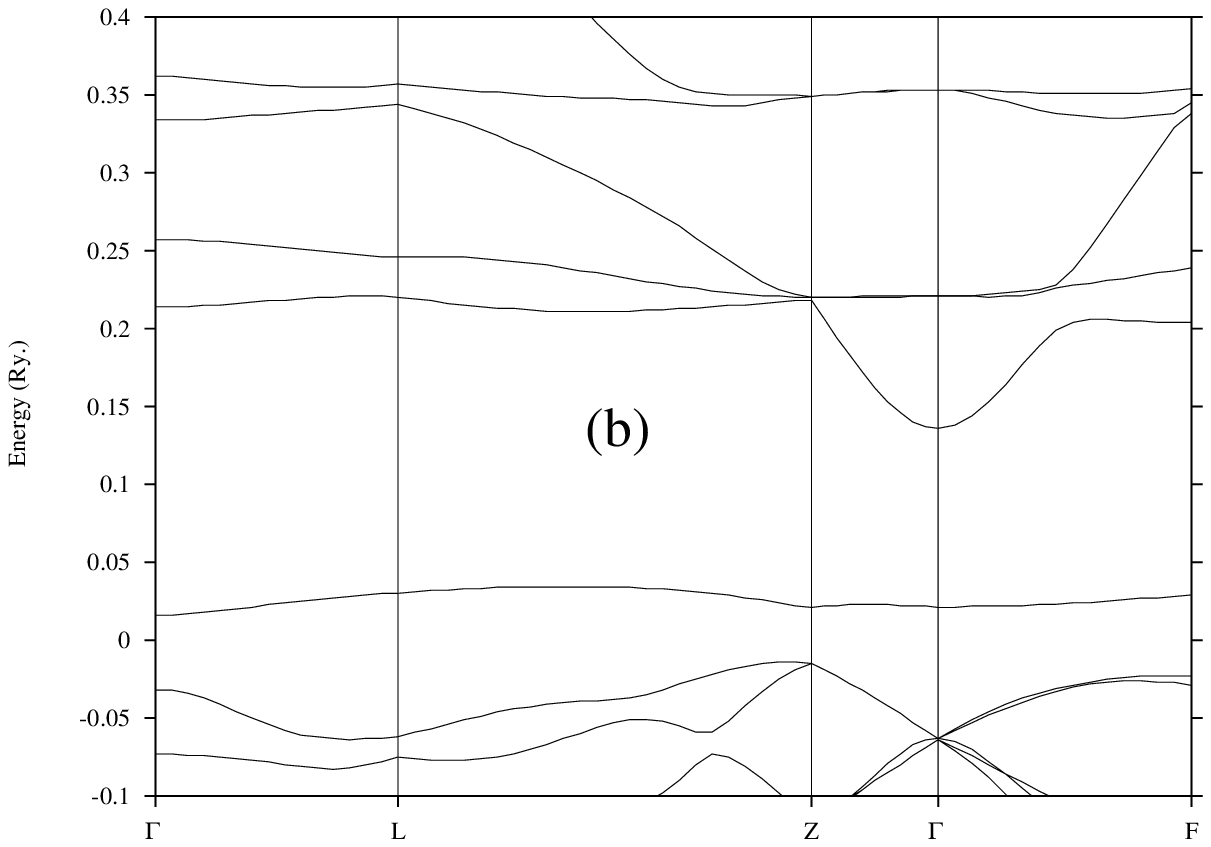,width=0.95\linewidth}} 
\caption{(a) LDA band structure without a rhombohedral distortion.
The bars show the relative $A_{1g}$ character of the 
correspoding states; (b) LDA+U bands for the same crystal structure.
 Note the bottom of the Fe($s)$ band at the $\Gamma $ point.} \label{cubic} 
\end{figure}
The corresponding band structure is shown on Fig.\ref{cubic}, where the
LDA+U, equally successful in predicting a band gap, forms a completely
different insulating state. There is no mechanism for unquenching the
orbital moment, apart from the spin-orbit coupling,
neglected in Ref. \cite{LDA+U} and other
LDA+U calculations. LDA+U starts from the straight LDA
bands, where the $t_{1g}$ bands are strongly mixed, the $A_{1g}$ (in
rhombohedral notations) orbital being slightly more occupied than each of
the two $E_g^{\prime }$ orbitals. When formulated in rotationally invariant
form\cite{LA}, LDA+U tends to apply the positive $U$ correction to less
occupied $E_g^{\prime }$ orbitals, making them less and less filled
(potential $U$ correction for more filled $A_{1g}$-orbital is negative) ,
and eventually splits off the $A_{1g}$ band, forming a gap between it and
the $E_g^{\prime }$ bands. Sufficiently large $U$ (we used the empirical
value of $U$=5.1 eV\cite{U5}, which gives a good value for the gap;
constrain LDA calculations \cite{anigun} yield a somewhat larger $U$=6.8 eV)
pushes the occupied $A_{1g}$ band down close to, and for $U$=6.8 eV right
into, the O($p)$-bands manifold, and the unoccupied $E_g^{\prime }$ bands up
above the bottom of the Fe($s)$ band. This seems to be in qualitative
agreement with the photoemission experiments, which show the top of the
valence band to be of the mixed O$2p$-Fe$3d$ character \cite{PES}, and with
the optical experiments, which indicate a weak absorption between 0.5 and
2.0 eV, assigned to the ($pd)-s$ transitions, and strong absorption edge at
2.4 eV due to transitions into the Fe($d)$ band\cite{opt}. In our
calculations, the minimal gap opens between the O$(p)-$Fe$(d)$ band and the
Fe$(s)$ band at 1.3 eV and transitions into the Fe$(d)$ band start at 2.2
eV.

LDA and the LDA+U bands refer to the cubic FeO (here and below we show
results of the ASA LMTO\cite{lmto47} calculations with the unit cell volume
256 bohr$^3$/FeO, close to the LDA equilibrium volume\cite{FeO}).
Since only 1 orbital, $A_{1g},$ is occupied (see Table \ref{occ-cub-2}),
there is no orbital moment. One can expect that including weak (a few mRy)
spin-orbit coupling will create a small orbital moment, but hardly one
comparable with the spin magnetization. It is interesting that the way
the gap opens in LDA+U in the cubic structure is very ``LDA-like'': 
The only way to open a gap in an effective one-electron approximation is to
split the $t_{2g}$ band by (magnetic, in this case) rhombohedral symmetry
and to occupy the $A_{1g}$ orbital for FeO and the two $E_g^{\prime }$
orbital in CoO. In fact, this is exactly what happened in the LDA
calculations of Dufek et al\cite{schwarz}, who used an exotic LDA
functional, which gave poor total energies but did open gaps in both
compounds.

\begin{figure}
\centerline{\psfig{file=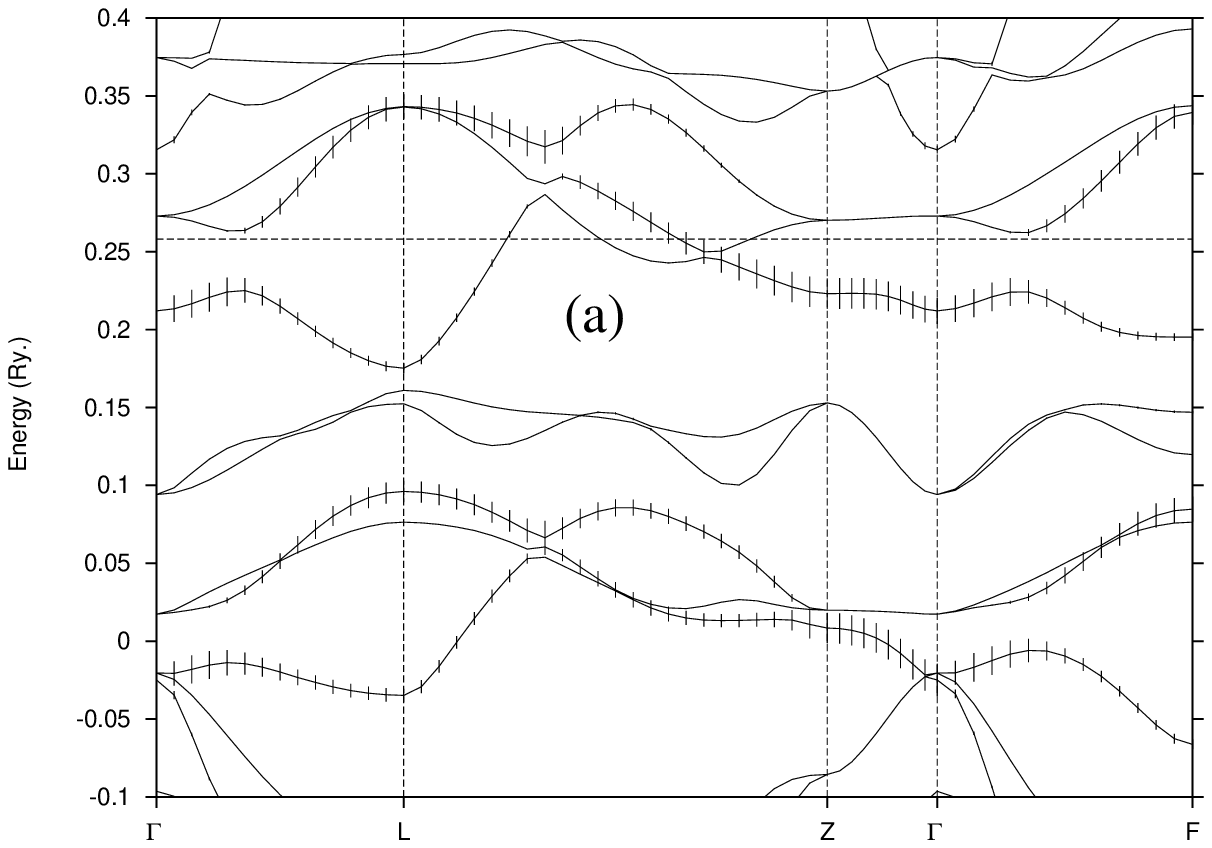,width=0.95\linewidth}} 
 \centerline{\psfig{file=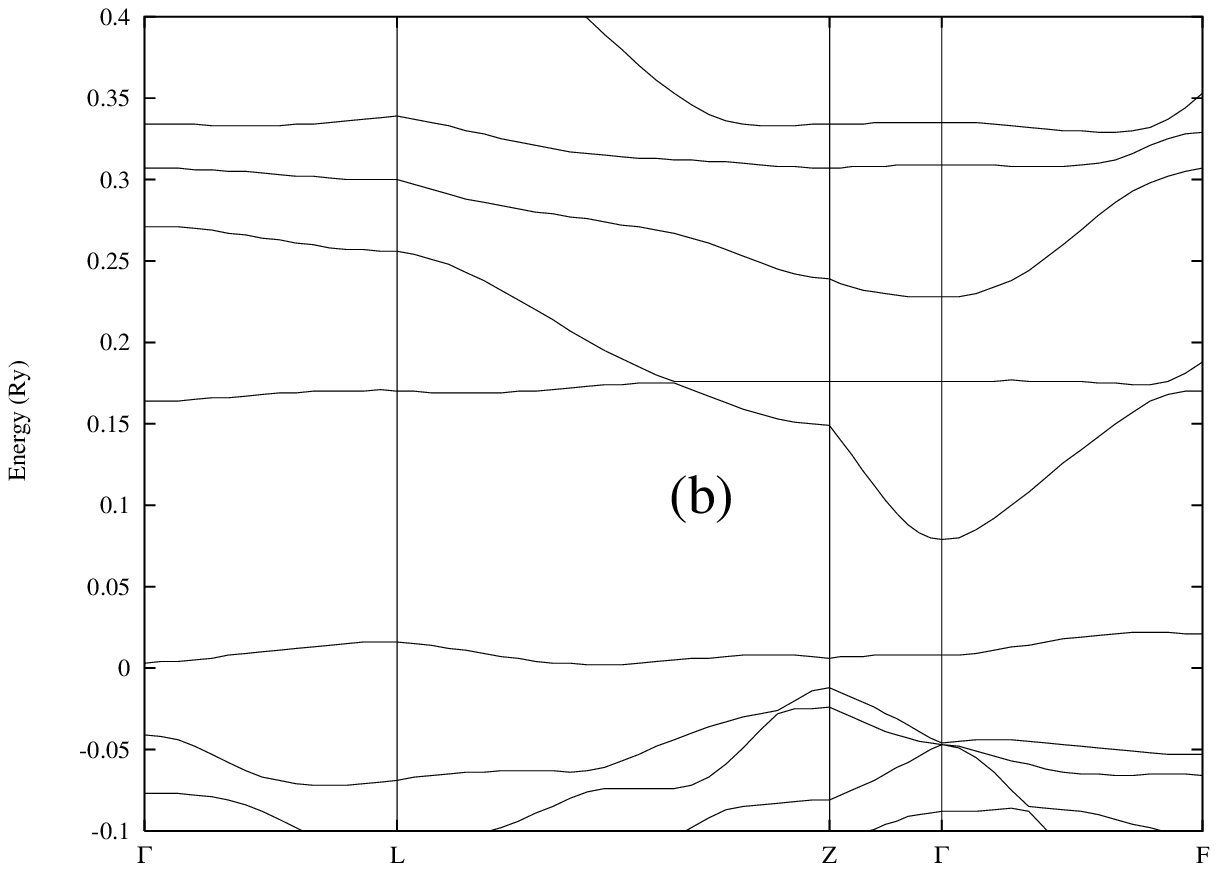,width=0.95\linewidth}} 
\caption{The same as Fig.1, but with the rhombohedral
distortion of 6\%.} \label{rhom} 
\end{figure}

Overall this seems to be a physically satisfactory description of the gap
opening. However, this is not as straightforward as one may think. To show
the problem, let us compare the calculations in the cubic structure with
those in the rhombohedrally distorted structure (Fig. \ref{rhom}; we used
the LDA equilibrium distortion, calculated in Refs. \onlinecite{FeO};
experimental distortion is smaller, but increases with the pressure). To
understand the result, one should keep in mind that dispersion of the $t_{2g}
$ bands is mainly due to direct $dd\sigma $ hopping between the like spins,
which can be easily verified in the LMTO-TB method
 by removing oxygen orbitals from
the basis set, or by looking directly on the corresponding elements of the
LMTO-TB Hamiltonian. We observe that the distortion increases the $t_{2g}$
bandwidth (due to decreased distance between the like-spin ions), but mainly
at the expense of the $E_g^{\prime }$-like bands. The dispersion of the $%
A_{1g}$ band decreases instead, thus leading to a decreased occupancy of the  $%
A_{1g}$-like state (because the whole $t_{2g}$ manifold is less than half
filled). One can see this by comparing the occupancy matrices in
the cubic (Table \ref{occ-cub-1}) and in the distorted (Table \ref
{occ-rhom-1}) structure on the first iteration, {\it i. e.}, before the
effect of $U$. While in the cubic structure the  $A_{1g}$ state was the
mostly occupied one (with the two $E_g^{\prime }$ states close next), in the
distorted structure the $E_g^{\prime }$ states are twice more occupied than
the $A_{1g}$ one.
The LDA+U code we used is rotationally invariant \cite{LA}, which means that
the program can apply the $U$ corrections to the states which are arbitrary
linear combinations of the $d$-orbitals with various $m$. The coefficients
of this combination are chosen self-consistently so that they optimize the
total energy including Coulomb one. In the case of the rhombohedrally
distorted FeO it is energetically unfavorable to occupy the $A_{1g}$
orbital, but not because of the Hubbard energy, but because of the
one-electron energy: the same reason why LDA deoccupied this orbital after
distortion. How does LDA+U handles this situation during the
self-consistency cycle? It is left with the only possibility to make a
linear combination of the $E_g^{\prime }$ orbitals with a possible admixture
of the $E_g^{\prime \prime }$ states, as much as the crystal field allows
that, to make it a separate band and to open a gap between this band and the
rest of the spin-minority $d$-bands (Table \ref{occ-rhom-2}). 

One may feel it unphysical that the method converges to qualitatively
different ground states with and without distortion. This fact suggests that
probably the correct solution should be intermediate between the two LDA+U
stationary points, a completely full and a completely empty  $A_{1g}$ state.
Indeed, a closer look at the LDA band structure of the Fig.\ref{rhom}a shows
that along the whole $\Gamma $-Z (111) direction the  $A_{1g}$ band lies
entirely below the Fermi level. In the other parts of the Brillouin zone
(points L and F) it is completely above the Fermi level. Thus, a $U$ matrix
acting on an unoccupied $A_{1g}$ orbital has to overcome the natural LDA
tendency along the  $\Gamma $-Z line, although it is in accord with the LDA
tendencies in the rest of the zone. 

From this consideration one can see that the problem with the LDA+U is that
the $U$-matrix is {\bf k}-independent: it is a matrix in the orbital space, $%
U_{mm^{\prime }}$, $-2\leq m\leq 2$, and this matrix does not know whether
the particular $d$-electron is in a bonding or antibonding interaction with
the neigboring sites. While this may be a good approximation for extremely
localized electrons, in such materials as FeO one cannot neglect the fact
that $d$-orbitals extend into neighboring sites. In the LMTO scheme the
tails of each orbital, penetrating into an atomic sphere of a neigboring
site, is re-expanded and, depending on the LMTO flavor (``representation''),
may or may not appear in the density matrix at the same site as its head\cite
{LMTO}. In the conventional LDA+U scheme, as well as in all other correction
schemes discussed above, only the heads of the LMT orbitals are subjects to
a correction. On the other hand,a more realistic correction would also apply
to that part of a $d$-orbital that penetrates into the neighboring spheres.
In other words, a physically correct scheme for implemented such a
correction should be essentially non-local. Hartree-Fock like schemes,
similar to the GW approximation\cite{GW}, should be able to reproduce the
correct physics. For instance, one can expect that the occupied band that
appears to be purely  $A_{1g}$ in the cubic case and purely $E_g^{\prime }$
in the distorted case, would have a mixed character, being more  $A_{1g}$%
-like close to the $\Gamma $-Z line. Correspondingly, the effective $U$%
-correction matrix would be different when applied at different {\bf k}%
-points. Another interesting alternative is offered by non-Kohn-Sham
versions of the Density Functional Theory, where a non-local (screened
Hartree-Fock like) contribution to the total energy is singled out together
with the non-interacting kinetic energy, and the rest is treated in an LDA.
Such schemes retain the good accuracy of the total energy
calculations in the LDA, and improve the excitation energies as well. It is
worth noting that underestimation of the hybridization tendencies inside the 
$t_{1g}$ band in LDA+U and other schemes is likely to lead to a wrong
energetics and spoil the agreement of structural properties, calculated in
LDA \cite{FeO}, with the experiment. More detailed discussion of structural
and magnetoelastic properties of FeO will be published elsewhere\cite{MRI}.

To summarize, we report conventional LDA calculations and
rotationally-invariant LDA+U calculations for antiferromagnetic FeO, both in
the cubic and in a rhombohedrally distorted structure. In both cases LDA+U
opens gaps, but these gaps are of completely different character in the two
structures, and also different from the gaps appearing in other
``corrected-LDA'' schemes, which, in turn, differ by their physical nature
from each other. We believe that this is a consequence of the local
character of the Mott-Hubbard correction in the conventional LDA+U method,
and neither straight LDA, nor ``corrected-LDA'' methods provide a proper
physical description of the gap formation. We argue that the ultimate method
must take into account the non-local character of the Coulomb repulsion.

We are thankful to O.K. Andersen, R.E. Cohen, D.I. Khomskii, A.I.
Liechtenstein, and W.E. Pickett
 for useful discussion. We acknowledge the hospitality of the
Max-Planck-Institut f\"{u}r Festk\"{o}rperforschung, where most of this work
was performed. One of us (VIA) was in part 
supported by Russian
Foundation for Fundamental Research (grant
RFFI 96-02-16167).

\begin{table}[tbp]
\caption{Orbitals and their occupations without rhombohedral distortion,
for $U=0$. Coordinates system corresponds to the rhombohedral symmetry:
z-axis is perpendicular to the ferromagnetic Fe planes, y-axis points
towards the nearest neighbor Fe with the like spin. 
The character of the orbitals is, in the order of the table,
$2\times E_g',\ 2\times E_g'',\ A_{1g}$ }
\label{occ-cub-1}
\begin{tabular}{l|ccccc}
 occupation & $xy$ & $yz$ & $3z^2-1$& $xz$&$x^2-y^2$ \\
\tableline 
.17 & -.03 & .03 & .00 & -.65 & .76\\ 
.17 & .75 & -.66 & .00 & -.03 & .03\\ 
.31 & .60 & .69 & .00 & .30 & .26\\
.33 & -.26 & -.30 & .00 & .69 & .60\\
.39 & .00 & .00 & 1.00 & .00 & .00
\end{tabular}
\end{table}
\begin{table}[tbp]
\caption{Orbitals and their occupations without rhombohedral distortion,
for $U=5.1$. The character of the orbitals is
$2\times E_g'',\ 2\times E_g',\ A_{1g}$}
\label{occ-cub-2}
\begin{tabular}{l|ccccc}
 occupation & $xy$ & $yz$ & $3z^2-1$& $xz$&$x^2-y^2$ \\
\tableline
     .04&    -.10  &  -.10&     .00   &  .68&     .72\\
     .04&    .72    &.68&     .00    &.10&    .10\\
     .10&     .53    &-.57&     .00 &   -.46&     .43\\
     .10&    -.43     &.46&     .00&    -.57&     .53\\
     .90&     .00   &  .00&   1.00&     .00&     .00
\end{tabular}
\end{table}

\begin{table}[tbp]
\caption{Orbitals and their occupations with 6\% rhombohedral distortion,
for $U=0$. The characters are $2\times E_g',\ A_{1g},\ 2\times E_g''$}
\label{occ-rhom-1}
\begin{tabular}{l|ccccc}
 occupation & $xy$ & $yz$ & $3z^2-1$& $xz$&$x^2-y^2$ \\ 
\tableline 
0.17 & -.55 & .84 & .00 & -.02 & .01\\
0.17 & -.01 & .02 & .00 & .83 & -.56\\
0.19 & .00 & .00 & 1.0 & .00 & .00\\
0.38 & .84 & .55 & .00 & .00 & -.00\\
0.39 & -.01 & -.01 & .01 & .56 & .83
\end{tabular}
\end{table}
\begin{table}[tbp]
\caption{Orbitals and their occupations with 6\% rhombohedral distortion,
for $U=5.1$. The characters are $A_{1g},\ E_g'',\ 2\times E_g',\ 
A_{1g},\ E_g''$}
\label{occ-rhom-2}
\begin{tabular}{l|ccccc}
 occupation & $xy$ & $yz$ & $3z^2-1$& $xz$&$x^2-y^2$ \\
\tableline
.04 & -.02 & .00 & .99 & .14 & .04\\
.04 & .96 & .28 & .02 & .01 & .01\\
.10 & -.28 & .96 & .00 & .02 & -.01\\
.10 & .01 & -.02 & -.09 & .83 & -.56\\
   .86&    -.01 &   .00&    -.12   &.55&     .83

\end{tabular}
\end{table}



\begin{references}
\bibitem{sic}  A. Svane and O. Gunnarsson, Phys. Rev. Lett. {\bf 64}, 1162
(1990); Z. Szotek, W.M. Temmerman, and H. Winter, Phys. Rev. B {\bf 64},
4029 (1993).

\bibitem{LDA+U}  V.I. Anisimov, J. Zaanen and O.K. Andersen, Phys. Rev. B 
{\bf 44}, 943 (1991);

\bibitem{norman}  M.R. Norman, Phys. Rev. B {\bf 44}, 1364 (1991); Int. J.
Quant. Chem. Symp. {\bf 25, }431 (1991).

\bibitem{coo}  CoO has tetragonal distortion, but no full-potential LDA
calculations for tetargonal phase were reported in the literature.

\bibitem{FeO}  D.G. Isaak, R.E. Cohen, M.J. Mehl, and D.J. Singh, Phys. Rev.
B {\bf 47}, 7720 (1993); D.M. Sherman and H.J.F. Jansen, Geophys. Lett. {\bf %
22}, 1001 (1995).

\bibitem{Good+Dan} J.B. Goodenough, Phys. Rev. B{\bf 171}, 465 (1966);
K.I. Kugel and D.I. Khomskii, Sov. Phys. Usp., {\bf 25}, 231 (1982).

\bibitem{zx}  Z.X. Shen {\it et al}, Phys. Rev. Lett. {\bf 64}, 2442 (1990);
Phys. Rev. B {\bf 42}, 1817 (1990); {\it ibid, }B {\bf 44}, 3604 (1991).

\bibitem{griffith}  J.S.Griffith,{\it The theory of transition metal ions},
Cambridge, 1971

\bibitem{LA}  A.I. Lichtenstein, J. Zaanen, V.I. Anisimov, Phys. Rev. B{\bf 52},
R5467 (1995)

\bibitem{U5}  J. Zaanen and G.A. Sawatzky, J. Solid State Chem. {\bf 88}, 8(1990)

\bibitem{anigun}  V. I. Anisimov and O. Gunnarson, Phys. Rev. B {\bf 43},
7570 (1991).

\bibitem{PES}  P.S. Bagus, C.R. Brundle, T.J. Chuang, K. Wandelt, Phys. Rev.
Lett. {\bf 39},1229 (1977)

\bibitem{opt}  I. Balberg, H.L. Pinch, J. Magn. Magn. Mater. {\bf 7}, 12 (1978)

\bibitem{schwarz}  P. Dufek, P. Blaha, and K. Schwarz, Phys. Rev. B {\bf 50}%
, 7279 (1994);

\bibitem{lmto47}  We used the LMTO-TB-ASA code {\it Stuttgart 4.7.}

\bibitem{LMTO}  O.K. Andersen and O. Jepsen, Phys. Rev. Lett., {\bf 53},
2571 (1984).

\bibitem{GW}  F. Aryasetiawan and O. Gunnarsson, Phys. Rev. Lett., {\bf 74},
3221 (1995).


\bibitem{MRI}  I.I. Mazin, R.E. Cohen, and D.G. Isaak, to be published.
\end{references}
\end{document}